\documentclass[12pt]{amsart}
\usepackage{epsfig,amssymb,amsfonts}
\usepackage[latin1]{inputenc} 
\usepackage{fancyheadings} 
\newcommand{\bds}[1]{\boldsymbol{#1}}
\def\half{\mbox{\small $\frac{1}{2}$}}
\def\VEV#1{\left\langle #1\right\rangle}

\def\fourth{\mbox{\small $\frac{1}{4}$}}
\def\p{{\mathfrak p}}                            
\def\O{{\mathcal O}}                            
\newcommand{\der}[2]{\dfrac{d #1}{d #2}}

\begin{document}

\begin{titlepage}
\begin{flushright}
     UPRF2000-01\\
     SWAT:253 \\
     February 2000 \\
\end{flushright}
\par \vskip 10mm
\begin{center}
{\Large \bf Understanding stochastic perturbation
 theory: toy models and statistical analysis\footnote{ Research
 supported by Italian MURST under contract 9702213582, by
 I.N.F.N. under {\sl i.s. PR11} and EU Grant EBR-FMRX-CT97-0122.}}
\end{center}
\par \vskip 2mm
\begin{center}
{\bf R.\ Alfieri $\,^a$,
F.\ Di Renzo$\,^a$,
E.\ Onofri}$\,^a$ \\[.5 em]
and\\
{\bf L.\ Scorzato}$\,^{a,b}$\\[.5 em]
$^a\,${\sl Dipartimento di Fisica, Universit\`a di Parma} \\
and {\sl INFN, Gruppo Collegato di Parma, Parma, Italy}\\[.5 em]
$^b\,${\sl Department of Physics, University of Wales Swansea, UK} \\[.5 em]
\vskip 2 mm
\end{center}
\par \vskip 2mm
\begin{center} { \large \bf Abstract} 
 \end{center}
\begin{quote}
{\sl The numerical stochastic perturbation method based on Parisi--Wu
quantisation is applied to a suite of simple models to test its
validity at high orders. Large deviations from normal distribution for
the basic estimators are systematically found in all cases (``Pepe
effect'').  As a consequence one should be very careful in estimating
statistical errors. We present some results obtained on Weingarten's
``pathological'' model where reliable results can be obtained by an
application of the {\sl bootstrap} method.  We also present some evidence
that in the far less trivial application to Lattice Gauge Theory a similar
problem should not arise at moderately high loops (up to $O(\alpha^{10})$). 
}
\end{quote}
\end{titlepage}

\section{Introduction}
In a series of papers \cite{DRMMOLatt94,DRMMO94,BDRMOPS99a} it has
been shown that the technique of stochastic quantisation introduced by
Parisi and Wu \cite{PaWu} can be implemented as a practical algorithm
which enables to reach unprecedented high orders in lattice
perturbation theory (e.g. $\alpha^8$ in the plaquette expectation
value in 4-D $SU(3)$ lattice gauge theory). Evaluating perturbative
expansion coefficients by a Monte Carlo technique opens the back--door
to a number of errors, both statistical and systematic, which should
be understood and, hopefully, kept under control. It has been shown in
Ref.\cite{DiRMO97} how systematic errors due to the finite volume can be
estimated, putting a limit on the perturbative order reachable on a
given lattice size. The aim of the present paper is to present a
rather detailed analysis of statistical errors, which present some
novel features with respect to the ordinary practice in lattice gauge
Monte Carlo.  This work was triggered by the observation made by
M. Pepe \cite{pepe96} 
of unexpected large discrepancies with respect to known values in the
perturbative coefficients of the non--linear $O(3)
\sigma-$model. Starting from this observation (``Pepe effect'') we
performed a systematic study of simple models with a small number of
degrees of freedom trying to trace the origin of the discrepancies and
to resolve them in a reliable way. The crucial fact that our analysis
has uncovered is the following: the statistical nature of the
processes which enter into the calculation of perturbative
coefficients is rapidly deviating from normality as we increase the
perturbative order, i.e. the distribution function of a typical
coefficient estimator is strongly non--Gaussian, exhibiting a large
skewness and a long tail; as a result very rare events give a
substantial contribution to the average. A simple minded statistical
analysis based on the assumption of normality may grossly fail to
identify the confidence intervals; some {\sl non--parametric}
statistical analysis, like the {\sl bootstrap} method, is necessary to
assess the statistical error and provide reliable confidence
intervals. This idea will be shown at work in the analysis of a
lattice toy model (Weingarten's ``pathological'' model
\cite{weingarten80}) which nevertheless presents many features of
interest.

In view of this analysis one should of course worry about the results
obtained in Lattice Gauge Theory (LGT). Our main conclusion with this
respect is that one is not going to jeopardize the picture we drew in
our previous works. As amazing as it can appear at first sight, the
application of the method to a by far less trivial model stands on a
by far firmer ground. We shall in fact show how the distribution
function of coefficient estimators for a typical LGT observable does
not exhibit the strongly non--Gaussian nature we find in simpler
models.

The content of the paper is organized as follows: in sec.2 we recall
the basis facts about Parisi-Wu stochastic technique applied to the
numerical calculation of perturbative coefficients in quantum field
theory on the lattice (hereafter the ``Parisi-WU process'' or
PW-process for short). In sec.3 we discuss the probability
distributions of the PW-process; we shall argue that the customary
asymptotic analysis of ``sum of identically distributed independent
random variables'' does not really help at high orders; in this case
we shall present numerical evidence showing that the distribution
functions still present large deviations from normality, in particular
a whole window where the density presents a power--law rather than
Gaussian behaviour. We present some detail on the algorithmic
implementation of the PW-process in Sec.4 and some numerical results
in Sec.5. A discussion of an alternate route is given in Sec.6, based
on Girsanov's formula. We then show (sec.7) how a bootstrap analysis
can be very effective in estimating confidence intervals.  Finally, in
sec. 8, some evidence is produced that both convergence time of the
processes and statistical errors are under control for LGT. We present
our conclusions in sec. 9. Some details on the bootstrap method and a
formal analysis of convergence of perturbative correlation functions
are given in appendix.

\section{Stochastic perturbation theory}
Starting from Parisi and Wu's pioneering paper, stochastic equations
have been used in various forms to investigate quantum
field--theoretical models, both perturbatively and
non--perturbatively. In particular a Langevin approach can be used as
a {\sl proposal} step subject to a Metropolis check to implement a
non--perturbative MonteCarlo for Lattice Gauge Theories. We are
concerned here with another approach which makes use of the Langevin
equation to calculate weak coupling expansions. The idea
\cite{DRMMOLatt94,DRMMO94}  is very simple: we start from the
Langevin equation (let us focus our attention on a simple scalar field
$\varphi$)
\begin{equation}\label{eq:langevin}
\frac{\partial\varphi(\bds{x},t)}{\partial t} 
= -\frac{\partial S}{\partial \varphi(\bds{x},t)} + \eta({\bds x},t)
\end{equation}
where $\eta({\bds x},t)$ is the standard white--noise generalized
process; assuming that the action $S$ is splitted into 
a free $S_0$ and an interaction part $g\,S_1$, we expand the 
process $\varphi$ into powers in the coupling constant 
\begin{equation}\label{eq:expansion}
\varphi({\bds x},t) = \sum_{n=0}^\infty g^n\,\varphi_n({\bds x}, t)\;.
\end{equation}
The Langevin equation is thus translated into a hierarchical system
of partial differential equations 
\begin{eqnarray}\label{eq:stoch}
\frac{\partial\varphi_0({\bds x},t)}{\partial t} &=&
 -\frac{\partial S_0}{\partial \varphi_0({\bds x},t)} + \eta(\bds
{x},t)\\
\frac{\partial\varphi_n({\bds x},t)}{\partial t} &=&
 -G_0^{-1}\,\varphi_n({\bds x},t) + D_n(\varphi_0,\ldots,\varphi_{n-1})\;,
\; {\rm for}\; n \ge 1
\end{eqnarray}
$G_0$ being the free propagator and $D_n$ representing source terms which
can be expressed in terms of higher functional derivatives of the
interaction term $S_1$. Notice that the source of randomness is
confined to the first (free--field) equation; the system can be
truncated at any order $n$ due to its peculiar structure ($D_n$
depends only on $\varphi_m$ with $m<n$).

The system can be used to generate a diagrammatic expansion, as Parisi
and Wu did for gauge theories, in terms of the free propagator $G_0$;
or, it can be studied numerically by simulating the white--noise
process, as we have discussed in Ref.\cite{DRMMOLatt94}.  Any given
observable $\O(\varphi)$ can be expanded in powers of the coupling
constant

\begin{equation}
\O(\varphi)\equiv \O\left(\sum_{n\ge 0} g^n\, \varphi_n\right) = \sum_{n\ge 0} g^n\, \O_n\left(\varphi_0,\ldots,\varphi_{n}\right)
\end{equation}
and its expectation value is given by
\begin{equation}\label{eq:observable}
\VEV{\O} = \VEV{\sum_{n\ge 0} g^n \, \O_n} = \sum_{n\ge 0} c_n(\O)\,g^n\;.
\end{equation}
The operator $\O_n$ is therefore an unbiased estimator of the 
$n-$th expansion coefficient of $\VEV{\O}$. 

It is rather straightforward to implement this idea in a practical
algorithm, once the theory has been formulated on a space--time
lattice. The application to gauge theory was presented in
Ref.\cite{DRMMO94} using the Langevin algorithm of Batrouni {\sl et
al\/}. Finite size errors where studied in a subsequent paper
\cite{DiRMO97}.

The key problem we want to discuss in the present paper consists in
finding a reliable way to estimate the statistical errors in the
measure of $\VEV{\O_n}$. Admittedly, it would be desirable to have
some analytic information on the nature of the multidimensional
coupled stochastic processes ~(\ref{eq:stoch}).  As a substitute we
choose to study some toy models where we can perform high statistics
calculations and compare the results with the exact coefficients.

\section{Toy models.}
We have extensively studied the application of numerical stochastic
perturbation theory to the following simple models: 
\begin{enumerate}
\item[$i)$] quartic random variable: $S(\varphi,g)=\half \varphi^2+\fourth g \varphi^4\;, \qquad\varphi\in
\mathbb{R}$.
\item[$ii)$] dipole random variable: $S(\varphi,g)=[1-\cos(g\,\varphi)]/g^2,\;\qquad \varphi\in (-\pi,\pi)$.
\item[$iii)$] Weingarten's ``pathological model'' \cite{weingarten80}:
\begin{gather*}
S = \half \sum_\ell \varphi_\ell^2 + \fourth g \sum_{\p}\, \varphi_{\p}\\
\varphi_\p = \prod_{\ell\in\p} \,\varphi_\ell\;, \qquad\varphi_\ell\in\mathbb{R},
\end{gather*}
where $\ell$ runs over links and $\p$ over plaquettes in a simple
$n$-dimensional cell of a cubic lattice. We consider $n=2,3,4$. 
\end{enumerate}
The integration measure is $\prod d\varphi \exp(-S)$ with the ordinary
Lebesgue measure $d\varphi$ over all degrees of freedom; the
expectation value of any field observable is then given by 
\begin{equation*}
\VEV{\O} = Z^{-1}\int \prod d\varphi\; \O[\varphi,g]\, \exp\{-S(\varphi,g)\}\;.
\end{equation*}

The calculation of the weak coupling expansion for the typical
observables can be easily performed, except for for the model {\sl
iii)} in four dimensions, where we have been unable to go beyond 
the $5^{th}$ order. We have for instance for the ``quartic'' integral
\begin{align*}
\VEV{\varphi^4} &= -\fourth \,\der{\log(Z(g))}{g} =
1-3\,g-24\,g^2 -297\,g^3 -4896\,g^4-100278\,g^4\\ & -2450304\,g^5 - 
69533397\,g^6 - 2247492096\,g^7 - 81528066378\,g^8 + O(g^9)\;,
\end{align*}
for the ``dipole'' integral 
\begin{align*}
\VEV{\cos(gx)} = 1-\frac{g^2}{2}-\frac{g^4}{8}+\frac{17g^6}{96}
-\frac{29g^8}{512} -\frac{251g^{10}}{61440} + O(g^{12})\;,
\end{align*}
and finally 
{\makeatletter\@setsize\small{10\p@}\ixpt\@ixpt
\begin{align*}
\VEV{\varphi_\p} &\propto \der{\log(Z_3(g))}{g}   \\
\small
&=-\frac{g}{4} - \frac{13\,g^3}{64} - \frac{103\,g^5}{256} -
  \frac{23797\,g^7}{16384} - \frac{2180461\,g^9}{262144} -
  \frac{72763141\,g^{11}}{1048576} - \frac{13342715521\,g^{13}}{16777216} + O(g^{14})\;
(n=2)\\
&= - \frac{g}{4} - \frac{17\,g^3}{64} - \frac{595\,g^5}{1024} - 
  \frac{34613\,g^7}{16384} - \frac{3059191\,g^9}{262144} - 
  \frac{388561373\,g^{11}}{4194304} - \frac{67903544647\,g^{13}}{67108864} + O(g^{14})\;
(n=3)\\
&= - \frac{g}{4} -  \frac{21\,g^3}{64} - \frac{411\,g^5}{512} +
 O(g^7)\; (n=4)
\end{align*}
} for Weingarten's model.  We have limited ourselves to models
involving real random variables which make it very simple to implement
the stochastic differential equation Eq.\eqref{eq:stoch}. To reach
high orders we used a symbolic language to build the right-hand side
({\sl i.e.\/} the terms $D_n(\varphi_0,\ldots,\varphi_{n-1})$). In the
numerical experiments $n$ is in the range $10$ to $16$.

For the sake of brevity we shall discuss our numerical experiments for
the third model, which is the nearest to lattice gauge theory, but the
qualitative aspects of the results are indistinguishable in the other
models. 

Of course one should be aware that Weingarten's model is rather
peculiar; its action is not bounded from below and it may make sense
only in Minkowski space. From the perturbative viewpoint, however, it is
perfectly admissible and for imaginary $g$ the integrals converge
absolutely. So this is an example where perturbation theory for a
model living in Minkowski space only can be computed by the stochastic
method even if the model does not allow for a  Euclidean formulation. 

\section{Algorithms}
Let us describe in some detail the numerical implementation of the
stochastic differential equations \eqref{eq:stoch}. The free field
$\varphi_0$ can be computed {\sl exactly} (as a discrete Markov chain) 
since it is a Ornstein-Uhlenbeck process \cite{wax}. For a time
increment $\tau$ we can write
\begin{equation}
\varphi_0(t+\tau) = e^{-\tau}\varphi_0(t) + \sqrt{1-\exp(-2\tau)}\,N(0,1)
\end{equation}
where $N(m,\sigma)$ is a normal deviate with mean $m$ and variance
$\sigma$.  By integrating over the time interval $[t,t+\tau] $ we get
\begin{eqnarray} \label{form_sol}
\varphi_n(t+\Delta t) =  e^{-\tau}\varphi_n(t) + \int_t^{t+\tau}
\exp(t'-t-\tau)\,D_n(\varphi_0(t'),\ldots,\varphi_{n-1}(t'))\,dt'
\end{eqnarray}
which can be approximated, for instance, by the trapezoidal rule. Due
to the peculiar nesting of the differential equations it is not
necessary to perform the usual ``predictor'' step to implement the
trapezoidal rule, which results in a faster algorithm.

The bias introduced by the approximation is of order $O(\tau^\nu)$,
with $\nu\approx 2$, which should be taken into account as it is usual
with the numerical implementation of Langevin equation. In this case
however we cannot apply a Metropolis step, as it can be done in the
non--perturbative case. The trapezoidal rule and the absence of bias
on the free field, however, conspire to keep the finite-$\tau$ error
to a small value.

The numerical experiments have been performed by running several
independent trajectories in parallel (typically $10^2$ to $10^4$); a
crude estimate of autocorrelation gives a value near to $\tau_{int}=\half$
which is the relaxation time of the Ornstein-Uhlenbeck process which
drives the whole system. Hence we measure the observable $\O_n$ every
$n_{skip}$ steps with $n_{skip}\,\tau \gtrsim 1$.

The pseudo-random numbers needed to generate the normal deviates are
produced through two different algorithms, the lagged--Fibonacci
algorithm described by Knuth \cite{knuth98} and Luscher's algorithm
\cite{luscher94}. They do not give appreciably different results in 
our experiments.                                             
                                                  
\section{Results}
We report the results of experiments carried out on the model {\sl
iii)} since it permits a graduation in dimensionality. Other models
have been studied in detail and they show the same general pattern.
The first three figures are organized as follows: the first plot is a
{\sl log-log} histogram of the raw data for a given observable
(i.e. $\O_{11}$); the third plot represents $\frac1T\int_0^T
\VEV{\O_n}\,dt$ averaged over 1000 independent histories. The middle
plot represents a blow up of the same average together with some
bootstrap sample which allow to estimate confidence intervals, as we
describe in some detail later on.  

\begin{figure}[t]
\mbox{\epsfig{figure=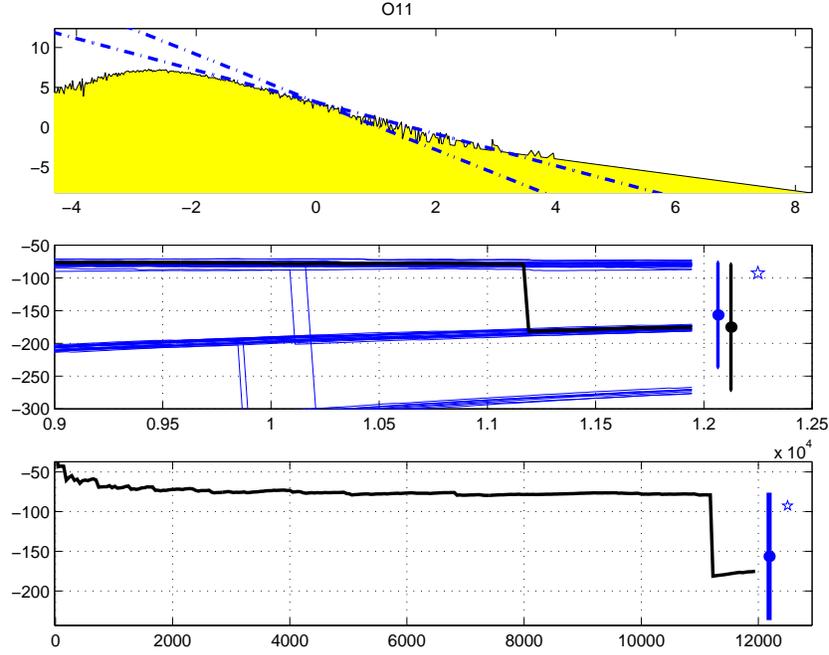,height=9.cm}}
\caption{An example of a very rare event in the $O_{11}$ history.}
\label{fig:pepeff}
\end{figure}

\begin{figure}[t]
\mbox{\epsfig{figure=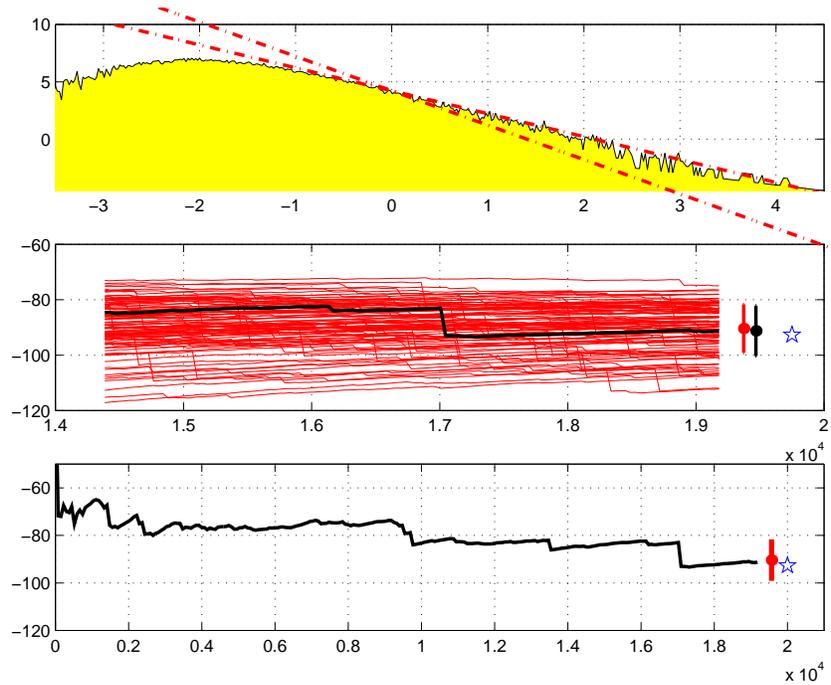,height=9.cm}}
\caption{A second $O_{11}$ experiment.}\label{fig:Oeleven}
\end{figure}

The main observation consists in the fact that the histograms
corresponding to high order coefficients deviate substantially from a
Gaussian distribution. We do not have a complete characterization of
the densities; however the study of these histograms sheds some light
on the problem. We observe a rather large window where the densities
are dominated by a power--law fall--off. Above a certain order,
typically 10-14, the power is approximately $x^{-2-\nu}$ with small
$\nu$, which means that there are large deviations from the average
which occur as very rare events but contribute to the average. An
example is found in Fig.~\ref{fig:pepeff} where the time average shows
a sharp kink at $t\approx 1.12e4$ and a relaxation thereof. In the
first picture the log-log--histogram is contrasted with a $x^{-2}$ and
$x^{-3}$ behaviour.  Several runs on the same model $iii_{3D})$ are
necessary to get a satisfactory estimate of the perturbative
coefficient (here $c_{11}=388561373/4194304\approx 92.64$). Another
experiment gives more regular histories (see Fig.~\ref{fig:Oeleven}).

At order 15 we encounter a similar pattern. We observe that the large
jumps at $t=1e4$ and $t=1.7e4$ both contribute to reach a value rather
close to the exact one. {\sl Were it not for the histograms, which, by
exhibiting a large $x^{-2-\nu}$ window, warn about very slow
convergence, one would be tempted to conclude that a plateau had been
reached well before $t\approx  10^4$}.

\begin{figure}[t]
\mbox{\epsfig{figure=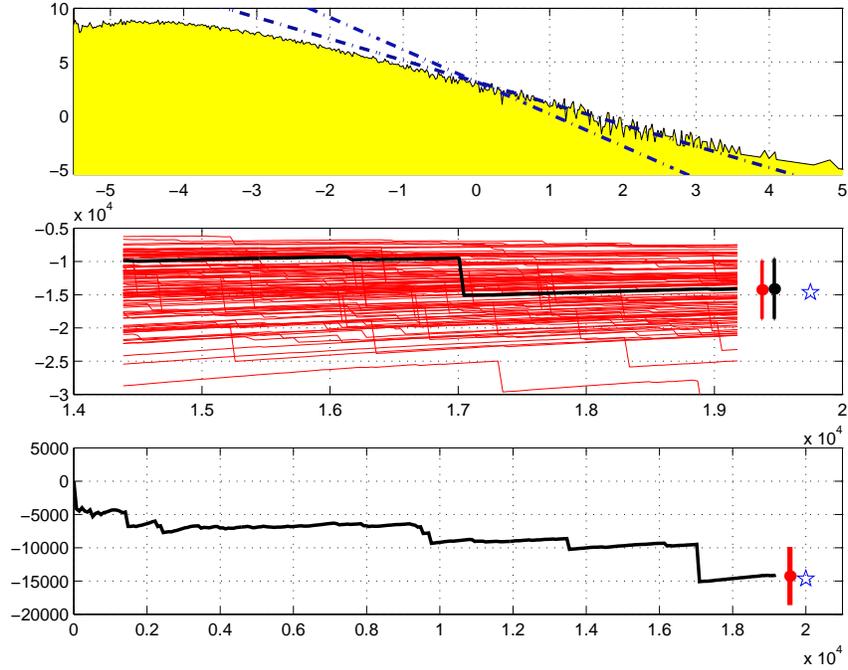,height=9.cm}}
\caption{A  $O_{15}$ experiment.}\label{fig:Ofifteen}
\end{figure}

On the contrary, low order coefficients are rather close to Gaussian
behaviour and the convergence is very fast (see Fig.~\ref{fig:Ofive}):

\begin{figure}[t]
\mbox{\epsfig{figure=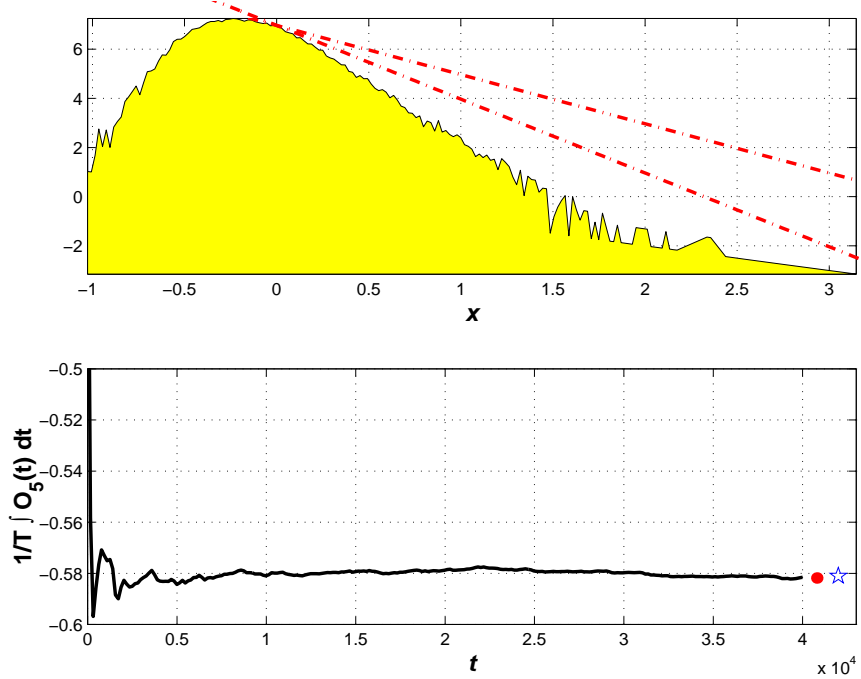,height=9.cm}}
\caption{A  $O_{5}$ experiment.}\label{fig:Ofive}
\end{figure}

We attempted an analytical characterization of the distribution
functions $P_n(\O_n<x)$ for the processes $\O_n$. The problem is
rather tricky and no explicit formula was reached. 
One can only show that if $\O$ is a generic correlation of fields, and
$P_n(\O_1, \ldots, \O_n,t)$ the distribution function of its first $n$
perturbative coefficients at a Langevin time $t$, then 
a limit distribution (for $t\rightarrow\infty$) always exists, and
that all its moments are finite. 
In fact this reduces to show that all correlation functions 
\begin{equation} \label{funzcorr}
\VEV{\prod_j \varphi_{p_j}(t) }
\end{equation}
are finite in the limit $t\rightarrow\infty$. Details can be found in
App.~\ref{sec:gigi}. 

Since all moments turn out to be finite, one is tempted to use general
theorems regarding the sum of independent identically distributed
random variables \cite{petrov95}. These could in principle make it
unnecessary to have a detailed knowledge of $P_n$, since we are
averaging over a large number of independent histories and the outcome
should be a Gaussian distribution with corrections which can be
parameterized and fitted to the data. According to Chebyshev and
Petrov
\begin{equation}
\int_{-\infty}^x [P^{(k)}_n(x) -G(x)] dx = 
\frac{e^{-x^2/2}}{\sqrt{2\pi}} 
(\frac{Q_1^{(k)}(x)}{\sqrt{n}} + \frac{Q_2^{(k)}(x)}{n} + \ldots)
\end{equation}
where the $Q_i(x)$ are polynomials, $G(u)$ is the normal distribution
and $k$ is the number of independent histories and the convergence is
even uniform on $x$. One has for instance
\begin{eqnarray*}
&& Q_1^{(k)}(x) = \lambda^{(k)}_3 \frac{(1-x^2)}{6}; 
\hskip10mm
Q_2^{(k)}(x) = \lambda^{(k)}_4 x(3-x^2); \\
&& \lambda^{(k)}_q = \frac{\VEV{(\phi^{(k)})^q}_c}{\sigma_k^q} ;
\hskip10mm 
\sigma_k^2 = \VEV{(\phi^{(k)})^2}. 
\end{eqnarray*}

Unfortunately the expansion is, at best, an asymptotic one and no
useful estimate on the error was obtained on the basis of these
formulas.  This fact has triggered our interest on non--parametric
methods of analysis which will be presented in
sec.~\ref{sec:bootstrap-analysis}.

 \begin{figure}[ht]\label{fig:girsa}
 \mbox{\epsfig{figure=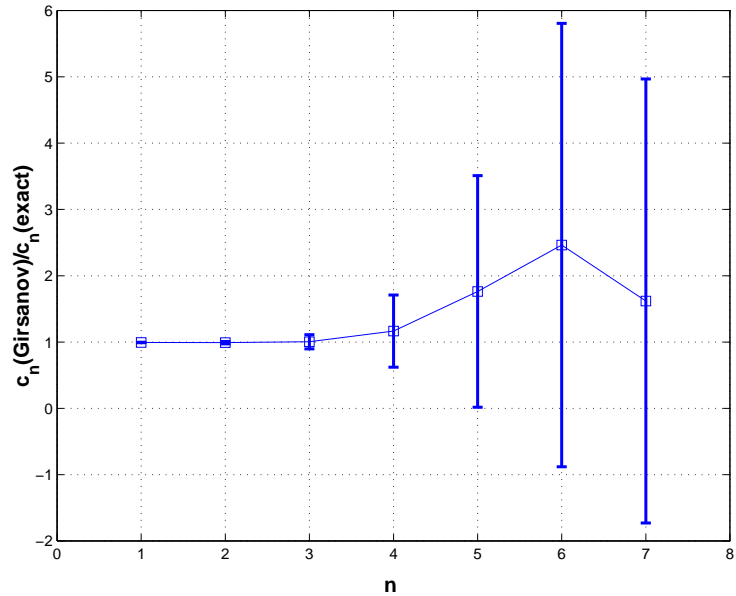,height=8.cm}}
 \caption{Model $i)$ by Girsanov's formula.}
 \end{figure}

 \begin{figure}[ht]\label{fig:girsabis}
 \mbox{\epsfig{figure=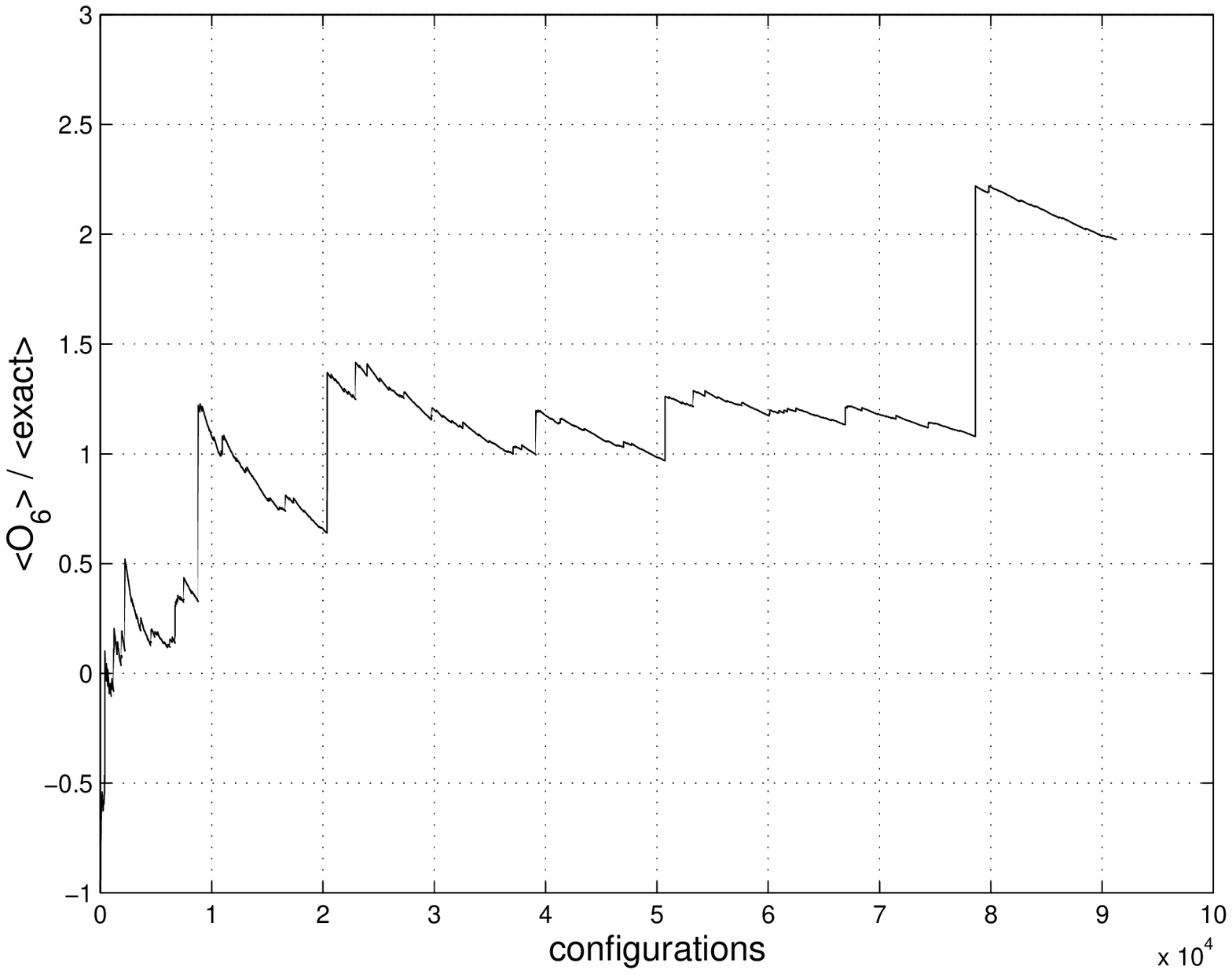,height=8.cm}}
 \caption{Large deviations in $\VEV{O_6}$.}
 \end{figure}

\section{Girsanov's formula}
The huge jumps in processes $\O_n$ at high $n$ could in principle be
caused by numerical inaccuracy, and for some time we suspected that
this could in fact happen. It was soon realized that the large
deviations are in fact needed to reach the right average in the long
runs, but it was nevertheless reassuring that an alternative method
with {\sl zero autocorrelation} and a totally different algorithm gave
the same pattern of configurations. As a matter of fact the method
turns out to be less efficient than the direct solution of the system
given by Eq.(\ref{eq:stoch}), but it is important in order to show
that the peculiar properties of the $\O_n$ histories are not an
algorithmic artifact.

The method applies Cameron-Martin-Girsanov formula (see
e.g.\cite{jona96}, \cite{gihman:72}, \S 3.12)
\begin{equation}
E\Bigl(\Phi\bigl(\bds x(T)\bigr)\;\Big\vert_w\Bigr) = E\Bigl(\Phi\bigl(\bds y(T)\bigr)\,\exp(\xi_T)\;\Big\vert_w\Bigr)
\end{equation}
where 
\begin{eqnarray*}
d\bds x(t) &=& -A\bds x \,dt + \bds b(\bds x,t) \,dt +d\bds w(t)\\
d\bds y(t) &=& -A \bds y\,dt + d\bds w(t)\;,\quad (y(0)=x(0))\\ \xi(T)
&=& \int_0^T \bds b(\bds y(\tau),\tau)\cdot d\bds w(\tau) -
\half\int_0^T \Vert \bds b(\bds y(\tau),\tau)\Vert ^2\,d\tau
\end{eqnarray*}
and $E\left(.\,\vert_w\right)$ denotes the average with respect to the
standard Brownian process $\bds{w}$.  We apply the formula like in
Ref.\cite{jona96}, hence $A$ is given by the free inverse propagator
and $\bds b(\bds x,t)$ is proportional to the coupling constant $g$;
therefore we can explicitly expand the ``exponential martingale''
$\xi(t)$ as a power series in $g$ and get an explicit characterization
of the expansion coefficients:
\begin{equation}
\VEV{\O_n} = \lim_{T\to\infty} E\left(H_n(I_1(T))\;I_2^\frac{n}{2}(T)/n!\right)
\end{equation}
where 
\begin{eqnarray*}
I_1(T)&=&\half\int_0^T\bds b(\bds y(\tau),\tau)\cdot d\bds w(\tau)\\
I_2(T)&=&\half\int_0^T\Vert \bds b(\bds y(\tau),\tau)\Vert d\tau
\end{eqnarray*}
and $H_n$ are Hermite polynomials. This representation of the
perturbative coefficients is completely independent on our expansion 
Eq.(\ref{eq:stoch}). It has been implemented as a numerical algorithm and
used to estimate the perturbative expansion for model $i)$. Each
iteration consists in selecting a normally distributed starting point
$x(0)$ and following the evolution up to a time $T$ where transient
effects are sufficiently damped; since $A=1$ in our model, $T\ge 1$ is
adequate. The samples are statistically independent, modulo the
quality of the random number generator. By monitoring the average of
$\xi(t)$, which should be exactly {\sl one}, we have a handle on the
accuracy of the algorithm (finite time step and statistics).

The application of Girsanov's formula gives a cross--check on the
existence of large deviations; the algorithm, however, is much more
cumbersome to implement on models other than the simple scalar field
and even in this latter case it turns out to be less efficient. 

\begin{figure}[ht]
\mbox{\epsfig{figure=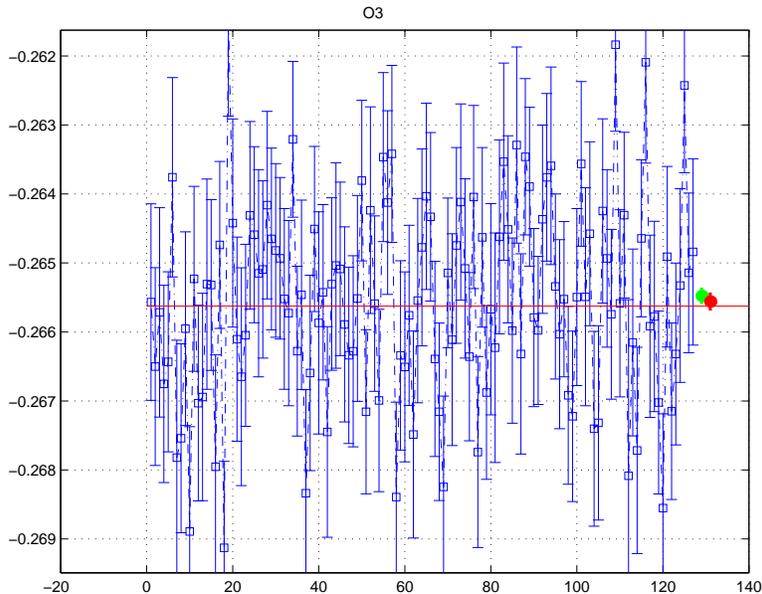,height=8.cm}}
\caption{Bootstrap analysis for $O_3$.}\label{figthree}
\end{figure}

\begin{figure}[ht]
\mbox{\epsfig{figure=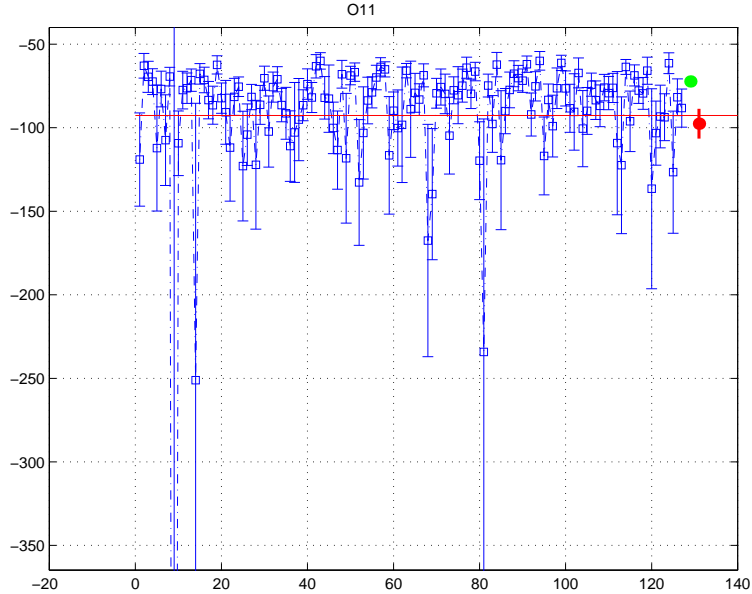,height=8.cm}}
\caption{Bootstrap analysis for $O_{11}$.}\label{figeleven}
\end{figure}

\section{Bootstrap analysis}\label{sec:bootstrap-analysis}
The consideration of a totally trivial random variable, a high power
of a simple normal deviate, is sufficient to convince ourselves that
the occurrence of large deviation is actually very natural. Consider a
standard Gaussian random variable $x$ and let $\O=x^n$. The
probability density for $\O$ contains a power--law prefactor which
dominates, for high $n$, over the exponential term. It is clear, by a
saddle point argument, that for $n\ge 5$ the average $\VEV{x^n}$ is
dominated by large deviations in the process; this appears to be a
common crux in all models we have considered. Having established that
the high order coefficients will suffer from large fluctuations, it is
important to use reliable tools to estimate the statistical
fluctuations. Since our estimators $\O_n$ will in general have
strongly non-Gaussian distribution functions, which moreover are only
known empirically through the numerical experiments, it appears a
natural choice to apply some {\sl non--parametric} analysis. We
present here the results obtained by applying the {\sl bootstrap}
method \cite{efron93}. Let us first consider experiments on the 3-D
Weingarten's model, restricted to an elementary cube (a total of 12
link variables, 6 plaquettes). We subdivide the data coming from
several runs at the same value $\tau=.01$ (measuring every 100 steps)
in bins of $N$ samples (say $N=1000$) and on each bin we perform $B$
bootstrap replicas (in this example $B=100$). Fig.\ref{figthree}
reports the result for $\O_3$. In this case the bootstrap analysis is
indistinguishable from a standard gaussian analysis, which means that
the distribution is not too far from normal. The standard gaussian
analysis is performed by taking a weighted mean of of averages in each
sample, {\sl i.e.}
\begin{eqnarray*}
\VEV{X}_{gauss} &=& \dfrac{\sum_j \VEV{x_j}/\sigma_j^2}	{\sum_j 1/\sigma_j^2}\\
\sigma^{-2} &=& \sum_j \sigma_j^{-2}
\end{eqnarray*}

A discrepancy between these values and the bootstrap's signals a
significant deviation from normality, like in Fig.\ref{figeleven}. 

\begin{figure}[ht]
\mbox{\epsfig{figure=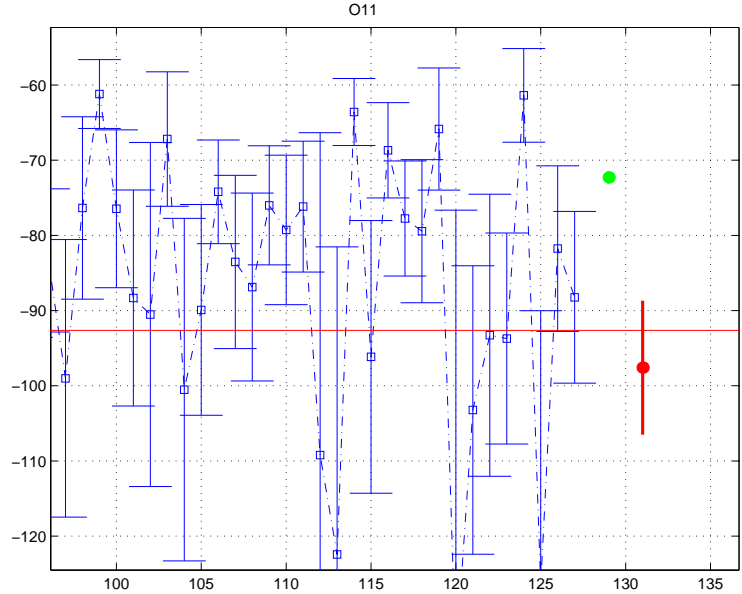,height=8.cm}}
\caption{A close up on the previous plot.}\label{figelevenbis}
\end{figure}

The next plot refers to the estimator $\O_{11}$. Here the deviation
from normal is relevant: it is clearly reflected in a strong
discrepancy between gaussian--like weighted mean and bootstrap
estimate. The bootstrap apparently gives a reliable estimate for the
confidence interval, as can be seen in the close up picture 
(Fig.~\ref{figelevenbis}).

\begin{figure}[ht]
\mbox{\epsfig{figure=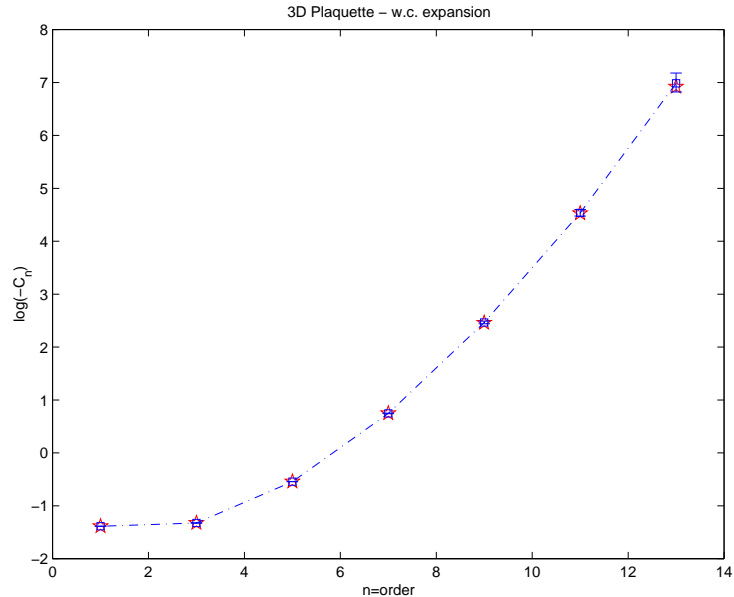,height=8.cm}}
\caption{Expansion coefficients for the plaquette variable in 3-D
``pathological'' model: $\star$=exact, dotted line = MonteCarlo.}
\label{fig:overall}
\end{figure}

Finally let us give a look at the overall pattern obtained in
3-D. Fig.~\ref{fig:overall} is remarkably similar to the one
obtained for the plaquette expansion coefficients in Wilson's 
4-D $SU(3)$ Lattice Gauge Theory. 

\section{Lattice Gauge Theories}

A good message from this paper is that the picture we just drew for
simple models does not spoil the confidence we have in Numerical
Stochastic Perturbation Theory for a field of real physical interest
like Lattice Gauge Theory. As amazing as it can appear at first sight,
this does not totally come as a surprise. In such a theory both the
huge number of degrees of freedom and their coupling make it less
likely applicable the simple argument of the previous section
referring to a high power of a normal variable. One should also keep
in mind that the gauge degrees of freedom make it possible to keep
the norms of the perturbative components of the field under control by
providing a restoring force which does not spoil the convergence of
the process.

\begin{figure}[t]
\mbox{\epsfig{figure=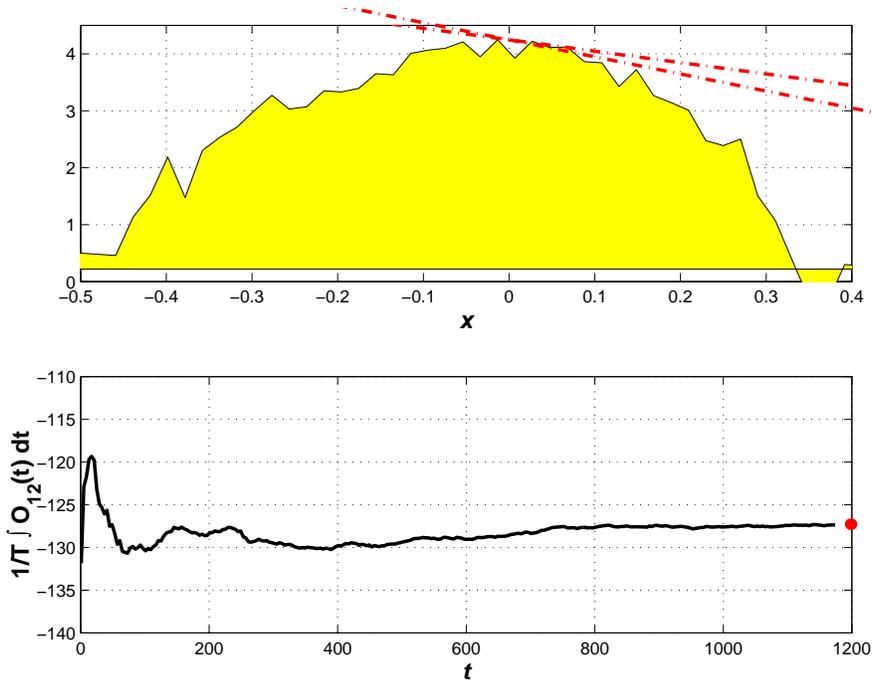,height=9.cm}}
\caption{The coefficient of $g^{12}$ for the $SU(3)$ plaquette on a
$4^4$ lattice.}
\label{fig:su3}
\end{figure}

As hard as formal arguments can be and since an exact characterization
of the stochastic processes is missing also for the simple models we
considered above, we take a pragmatic attitude. The lesson we can
learn from simple models is that monitoring frequencies histograms is
a good tool in order to assess huge deviations from normality. By
inspecting histograms from Lattice Gauge Theory simulations one can
convince oneself that in this case convergence is much more
reliable. This of course turns out to be consistent with the bootstrap
and standard error analysis giving the same results. For instance, the
plot in fig.~\ref{fig:su3} refers to a $SU(3)$ plaquette high order
perturbative coefficient on a small lattice: the process is manifestly
safe from ``Pepe-effect''.

An independent hint for the reliability of Numerical Stochastic 
Perturbation Theory for Lattice Gauge Theory comes from \cite{alfaten}. 
A couple of new perturbative orders in the expansion of the 
plaquette in 4-D $SU(3)$ (which means reaching $\alpha^{10}$) are shown 
to be fully consistent with both the expected renormalon behaviour and 
with finite size effects on top of that. A more organic report on the 
status of the art concerning the application of the stochastic method 
to LGT will be given in \cite{LGTreport}.

\section{Conclusions}
We have performed a series of simulations to test the numerical
stochastic perturbation algorithm previously introduced in the context
of Lattice Gauge Theories. The emergence of large non--Gaussian
fluctuations in the statistical estimators for high order coefficients
was observed in all the simple models we considered with a similar
pattern. The study of histograms, giving an estimate of the
distribution functions for these estimators, exhibits a large window
where exponential fall--off is masked by an approximately power--like
behaviour; this fact entails that in any finite run there exist large
deviations which are very rare but contribute to the average on the
same foot with more frequent events. The lesson we draw from this
experiment is twofold: it is necessary to monitor the histograms of
the estimators in order to assess the deviation from normality; in
case of large deviations, a reliable estimate of statistical error
should be obtained by non--parametric methods, such as the {\sl
bootstrap}.  These problems do not appear to plague the application of
the method to Lattice Gauge Theories. The same viewpoint presented for
simple models results in histograms hinting at good convergence
properties, fully consistent with all our previous experience in this
field.

\appendix

\section*{Acknowledgments}
Many people contributed ideas which finally merged in this paper. We
would like to thank {\sl A. Sokal}, who before anybody else raised
some doubts about the kind of convergence of the PW-process as we
implement it and stimulated a deeper analysis, {\sl P. Butera and
G. Marchesini} for their continuous interest and support, {\sl
M.Pepe}, for the original observation which triggered our analysis,
{\sl G.Burgio}, for many interesting discussions. E.O. would like to
warmly thank CERN's Theory Division for hospitality while part of this
work was performed.

\section{Correlation functions at high Langevin time.}\label{sec:gigi}

To simplify the argument, let us consider the model $(i)$ with a single
degree of freedom (it may be generalized also to realistic
systems). First of all we observe that any correlation function of
free fields $\varphi_{0}(t)$ may be exactly computed, in fact
\begin{equation} \label{freefields}
\VEV{\varphi_{0}(t)^{2m}} =
(2m-1)!!\, (1-e^{-2t})^m
\end{equation}
converges exponentially to a limit.  Then one can proceed by
induction: it is sufficient to show that the correlation function
Eq.(\ref{funzcorr}), which has a total perturbative order $p_T=\sum_j
p_j$, may be written as a finite sum of correlation functions which
have a total perturbative order strictly lower than $p_T$, plus
correlation functions which have a total perturbative order equal to
$p_T$ but with less free fields, plus exponentially damped terms.

To this end it is useful to re-write the formal solution given by
Eq.(\ref{form_sol}) in discretized time ($t=N\tau$). The idea is to
separate the solution into the memory terms, representing the last
$N-1$ steps, and the most recent contribution from lower orders and
noise.
\begin{eqnarray} \label{riscrittura}
\varphi_{0}(t) &=& e^{-\tau} \varphi_{0}(t-\tau) +
\sqrt{\tau}\eta(t)\\
\varphi_{j}(t) &=& e^{-\tau} \varphi_{j}(t-\tau) +
\tau f^{(j)}(t) \nonumber 
\end{eqnarray} 
where 
\begin{equation*}
f^{(j)}(t)=-\sum_{i=0}^{n-1}\sum_{j=0}^{i}
\varphi_{j}(t)\, \varphi_{i-j}(t)\,\varphi_{n-1-i}(t)
\end{equation*}	
and $\eta(t)$ are independent gaussian variables with mean zero and
variance $2$. We now should put Eq.(\ref{riscrittura}) inside
Eq.(\ref{funzcorr}). Let's do it for a correlation function of two
fields (the extension to a general correlation function will be
considered later). For any perturbative order $i, j > 0$ we have:
\begin{eqnarray*}
\VEV{   \varphi_{i}(N\tau)\,   \varphi_{j}(N\tau)  } &=&
e^{-2\tau}
\VEV{ \varphi_{i}(N\tau-\tau)\,\varphi_{j}(N\tau-\tau) } + \\
&&+ \tau e^{-\tau} 
\VEV{ \varphi_{i}(N\tau-\tau)\, f^{(j)}(N\tau) } \\
&&+ \tau e^{-\tau} 
\VEV{ \varphi_{j}(N\tau-\tau)\, f^{(i)}(N\tau) } \\
&&+ \tau^2 \VEV{ f^{(i)}(N\tau)\, f^{(j)}(N\tau) }
\end{eqnarray*}
Let us use the general solution of the recursive formula:
\begin{equation*}
y_l  = \alpha_l\, y_{l-1} + \beta_l
\end{equation*}
which is given by
\begin{equation*}
y_N = \left(\prod_{m=1}^N \alpha_m\right)\, y_0 
+ \sum_{n=1}^N \left(\prod_{m=n+1}^N \alpha_m\right)\,\beta_n
\end{equation*}

In our case we simply have $\alpha_m = e^{-2\tau}$, 
independent on $m$,
while $\beta_m$ is composed of a linear ($B_m$) and a quadratic
($C_m$) part in $\tau$:
\begin{eqnarray*}
\beta_m &=& \tau 
 \,e^{-\tau} \, B_m(\tau) + \tau^2 \, C_m\\
B_m(\tau)&=&
\VEV{ \varphi_{i}(N\tau-\tau)\, f^{(j)}(N\tau) } +
\VEV{ \varphi_{j}(N\tau-\tau)\, f^{(i)}(N\tau) } \\
C_m&=&\VEV{ f^{(i)}(N\tau)\, f^{(j)}(N\tau) }.
\end{eqnarray*}

As a result:
\begin{eqnarray} \label{dim0_18}
&&\VEV{   \varphi_{i}(N\tau)\,   \varphi_{j}(N\tau)   }   =
e^{-2 N\tau} \VEV{ \varphi_{i}(0)\, \varphi_{j}(0) } + \\
&+&e^{-2 N\tau}
\left[\sum_{n=1}^N  e^{2 n\tau}
(\tau e^{-\tau} B_n(\tau) +\tau^2 C_n )\right] \nonumber \\
\end{eqnarray}

We observe that $B_n(\tau)$ is a correlation function of total perturbative
order strictly lower than $i+j$. By inductive hypothesis we know that
 $B_n(\tau)$ has a finite limit for $\tau\rightarrow 0$ (at $t=N\tau$
fixed), let's say
\begin{equation} \label{ipind1}
B_N(\tau) = B_{t} + O(\tau),
\end{equation}
and that $B_{t}$ has a finite limit for $t\rightarrow\infty$. Let us
parameterize the remaining dependence on $t$ as:
\begin{equation} \label{ipind2}
B_t = B_{\infty} +  a_1 e^{-2t}\, t^p+
\sum_{q_j > 2} a_j e^{-q_j t}
\end{equation}
From Eq.(\ref{freefields}) it follows that the dependence on $t^p$ is
absent in correlations of free fields, but such correction factors can
appear at higher orders.  By inserting Eqs.(\ref{ipind1},\ref{ipind2})
in Eq.(\ref{dim0_18}) the geometric series can be re-summed:

\begin{eqnarray*}
&&\VEV{   \varphi_{i}(N\tau)   \varphi_{j}(N\tau)   }   =
e^{-2N\tau}
\VEV{ \varphi_{i}(0)\varphi_{j}(0) } + \\
&+&e^{-2N\tau}
\left[
\left(  \tau \, B_t + O(\tau^2)  \right)
\left( \frac{1-e^{2 (N+1)\tau}}{1-e^{2 \tau}}\right) + O(\tau)
\right]
\end{eqnarray*}
If we take first the limit $\tau\rightarrow 0$ at $t=N\tau$ fixed
and then $t\rightarrow\infty$, we find
\begin{equation*}
\lim_{t\rightarrow\infty}
\VEV{  \varphi_{i}(t) \varphi_{j}(t)} =
\tfrac{1}{2}  B_{\infty}.
\end{equation*}

Up to now, we have not considered the case of correlation functions
with some free fields together with higher order fields. In this
case the argument runs almost in the same way, but $\tau f^{(j)}$
is substituted by $\sqrt{\tau} \eta$:
\begin{eqnarray*}
\VEV{   \varphi_{0}(N\tau)   \varphi_{0}(N\tau) (\ldots) } &=&
e^{-2\tau}
\VEV{ \varphi_{0}(N\tau-\tau)\varphi_{0}(N\tau-\tau) (\ldots)} + \\
&&+ 2 \sqrt{\tau} e^{-\tau} 
 \VEV{ \varphi_{0}(N\tau-\tau)\, \eta(N\tau)\, (\ldots)} \\
&&+ \tau \VEV{ \eta(N\tau)\, \eta(N\tau)\, (\ldots)}
\end{eqnarray*}
Only terms with an even number of $\eta$ at time $N\tau$  contributes. 
In this case the inductive hypothesis must
be applied to correlation functions with the same total perturbative
order but with a lower number of free fields. 

The same argument can be applied to a general correlation function,
obtaining in the limit $t\rightarrow\infty$ 
\begin{eqnarray} \label{miaregola}
\VEV{ \prod_{l=0}^{p} \varphi_{l}^{n_l} } 
&=&
\frac{1}{\sum_{l=0}^{p}n_l}
\left[n_0(n_0-1)
\VEV{ \prod_{l=0}^{p} \varphi_{l}^{n_l}\,
\widehat{(\varphi_{0})^2}} -\right. \\
& &
\left.\sum_{m=1}^{p} n_m \sum_{i=0}^{m-1} \sum_{j=0}^{i} 
\VEV{ \prod_{l=0}^{p}\varphi_{l}^{n_l}\,
\widehat{\varphi_{m}}\,
 \varphi_{j}\,\varphi_{i-j}\, \varphi_{m-1-i}}\right] \nonumber
\end{eqnarray}
where $p$ is the maximum perturbative order present, and $\widehat{\varphi}$
means that a factor $\varphi$ should be dropped in the expression.

Induction allows us to conclude that all correlation functions reach a
finite limit.

If one goes through the previous argument by keeping the first
correction in Eq.(\ref{ipind2}) it is possible to show that
convergence to equilibrium is dominated by an exponential factor which
is at least $e^{-2t}$. However, even if in correlations of free fields
the dependence on $t$ has exactly an exponential form (or sum of
exponentials), in general (at higher orders) one must expect power
corrections like $t^\alpha \, e^{-2t}$. The precise determination of
them is tedious and beyond our scope.

The formula given in Eq.(\ref{miaregola}) is useful also because it
allows to compute not only the mean values of observables but also
some high moments of their distributions, at least for the model
($i$).  In this way we have been able to show that the large
fluctuations were neither an artifact of the numeric simulation nor the
effect of a slow convergence to equilibrium.

\section{The bootstrap algorithm} 

The idea of bootstrap is very simple. Let $X$ be a random variable and
$x_1, x_2, \ldots, x_N$ an $N-$tuple of values generated by a physical
process, by the stock market, by your Monte Carlo, whichever the case
you are currently studying.  In the absence of any a priori information
on the distribution function for $X$, one makes the most conservative
assumption, namely one adopts as distribution function the {\sl
empirical distribution} with density
\begin{equation*}
\rho_{boot}^N(x) = \frac1N \sum_{j=1}^N \delta(x-x_j)\;.
\end{equation*}
One then uses $\rho_{boot}$ to generate other $N$-tuples; all values
$x_j$, $(j=1,\ldots,N)$ being equally probable. Having produced $B$
such $N$-tuples one can estimate various statistics of interest, like
mean, median,  quartiles etc. For instance the definition of
standard deviation is simply 

\begin{equation}
\Delta x = \sqrt{\frac1{B-1} \sum_{k=1}^B \left( \VEV{x}_k - \VEV{x}\right)^2}
\end{equation}
where $\VEV{x}_k$ is the average computed on the $k$-th $N$-tuple and
$\VEV{x}$ is their mean value. Recommended values for $B$ are in the
range $25\lesssim B \lesssim 200$. It is obvious that no improvement can
be obtained on mean values; the virtue of the method should be
found in the easy estimation of standard errors which do not rely on
the assumption of normality. This does not mean of course that one
should not take care of autocorrelation problems or other sources of
statistical inaccuracies. 

The method is implemented very easily in any language. A one--liner to
produce a new $N$-tuple in {\tt matlab} is the following (if the
vector $X$ contains the original data)
\begin{quotation}
{\sf    XB = X(ceil(N * rand(N,1)));}
\end{quotation}
We refer to Refs.\cite{efron93,davison97} for a thorough discussion
of the method. 


\end{document}